\documentclass[12pt,thmsa]{article}

\usepackage{cite}

\usepackage{amsmath}

\begin{document}

\title{Nonadiabatic Calculation of Dipole Moments}
\author{Francisco M. Fern\'andez and Juli\'an Echave \\
INIFTA (UNLP, CCT La Plata-CONICET), \\
Diag. 113 y 64 (S/N), Sucursal 4, Casilla de Correo 16, \\
1900 La Plata, Argentina \\
e--mail: fernande@quimica.unlp.edu.ar, jechave@inifta.unlp.edu.ar}
\maketitle

\begin{abstract}
We review some of the few available nonadiabatic calculations of
dipole moments. We show that those carried out in a laboratory--fixed
set of coordinate axes are bound to fail and discuss the more reasonable
ones in a molecule--fixed reference frame. For completeness we also
describe the standard Born--Oppenheimer calculations of dipole moments.
We briefly address the experimental estimation of dipole moments from
the Stark shift of spectral lines and argue that it does not provide such
property but a sort of energy--weighted average of dipole transition
moments.
\end{abstract}

\section{Introduction\label{sec:intro}}

Many properties of molecular aggregates are attributed to the distribution
of charges in the constituent individual molecules or what is commonly
called the molecular dipole moment. This molecular property is also
responsible of the most salient features of the absorption and emission
spectra of molecules\cite{GST53,TS55,R56}. Molecular dipole moments can be
obtained, for example, from microwave spectra\cite{GST53,TS55,R56}, or
calculated from first principles\cite{SO96}.

The purpose of this review is to outline the calculation of molecular
dipole moments by means of quantum--mechanical approaches and compare such
theoretical results with corresponding experimental measurements. It is not
expected to be exhaustive because of the enormous number of worthy works on
the subject. However, we hope to succeed in giving an idea of the
difficulties encountered in such endeavour.

In Sec.~\ref{sec:Hamiltonian} we consider the nonrelativistic molecular
Hamiltonian and discuss the separation of the center of mass in detail. We
derive general expressions that may be useful for most of the various
nonadiabatic approaches that appear in the current literature. In Sec.~\ref
{sec:symmetry} we outline the dynamical symmetry of the molecular
Hamiltonian and some of the properties of the molecular stationary states.
In Sec.~\ref{sec:H-F} we discuss the well--known Hellmann--Feynman theorem
for variational wavefunctions because it is relevant to the calculation of
dipole moments. Although we are mainly interested in nonadiabatic
calculations of dipole moments, in Sec.~\ref{sec:BO} we consider the
separation of the electronic and nuclear motions that leads to the
Born--Oppenheimer and adiabatic approaches to the calculation of molecular
properties. For completeness, in Sec.~\ref{sec:Ext_field} we outline the
interaction between a system of point charges and an external electric field
with the purpose of defining dipole moment, polarizabilities and other
molecular properties. In Sec.~\ref{sec:exp} we briefly describe the models
used to derive the dipole moment from the Stark effect in rotational
spectra. They will be useful for comparison between experimental
measurements and theoretical calculations. In Sec.~\ref{sec:DM_BO} we
briefly review the theoretical calculation of molecular dipole moments under
the Born--Oppenheimer approximation. In Sec.~\ref{sec:DM_NBO} we discuss
some of the existing nonadiabatic calculations of dipole moments which is
the main topic of this review. In Sec.~\ref{sec:Mol-Fixed} we briefly
describe the more rigorous nonadiabatic calculations of dipole moments in
the molecule--fixed reference frame. For simplicity we restrict ourselves to
diatomic molecules because they appear to be the only ones accessible to
current nonadiabatic approaches. In Sec.~\ref{sec:PT} we outline the
application of perturbation theory to the Schr\"{o}dinger equation for a
molecule in an external electric field with the purpose of discussing a more
rigorous connection between the measured Stark shift and the molecular
dipole moment. Finally, in Sec.~\ref{sec:conclusions} we summarize the main
conclusions of this review.

\section{The molecular Hamiltonian}

\label{sec:Hamiltonian}

In this section we consider the nonrelativistic molecular Hamiltonian as a
system of $N$ charged point particles with only Coulomb interactions
\begin{eqnarray}
\hat{H} &=&\hat{T}+\hat{V},  \nonumber \\
\hat{T} &=&\sum_{i=1}^{N}\frac{\hat{p}_{i}^{2}}{2m_{i}},  \nonumber \\
V &=&\frac{1}{4\pi \epsilon _{0}}\sum_{i=1}^{N-1}\sum_{j=i+1}^{N}\frac{%
q_{i}q_{j}}{r_{ij}}  \label{eq:H=T+V_full}
\end{eqnarray}
In this expression $m_{i}$ is the mass of particle $i$, $q_{i}=-e$ or $%
q_{i}=Z_{i}e$ are the charges of either an electron or nucleus,
respectively, and $r_{ij}=|\mathbf{r}_{i}-\mathbf{r}_{j}|$ is the distance
between particles $i$ and $j$ located at the points $\mathbf{r}_{i}$ and $%
\mathbf{r}_{j}$, respectively, from the origin of the laboratory coordinate
system. In the coordinate representation $\mathbf{\hat{p}}_{i}=-i\hbar
\nabla _{i}$.

Since the Coulomb potential $V$ is invariant under translations $\hat{U}(%
\mathbf{a})\mathbf{r}_{i}\hat{U}(\mathbf{a})^{\dagger }=\mathbf{r}_{i}+%
\mathbf{a}$ ($\hat{U}(\mathbf{a})V\hat{U}(\mathbf{a})^{\dagger }=V$), then
the eigenfunctions of the translation--invariant Hamiltonian operator (\ref
{eq:H=T+V_full}) are not square integrable. For that reason we first
separate the motion of the center of mass by means of a linear coordinate
transformation
\begin{equation}
\mathbf{r}_{j}^{\prime }=\sum_{i}t_{ji}\mathbf{r}_{i}  \label{eq:r'->r}
\end{equation}
that leads to
\begin{equation}
\nabla _{j}=\sum_{i}t_{ij}\nabla _{i}^{\prime }  \label{eq:Nabla->Nabla'}
\end{equation}
and
\begin{equation}
\sum_{i}\frac{1}{m_{i}}\nabla _{i}^{2}=\sum_{i}\frac{t_{1i}^{2}}{m_{i}}%
\nabla _{1}^{\prime 2}+2\sum_{i}\sum_{j>1}\frac{t_{ji}t_{1i}}{m_{i}}\nabla
_{1}^{\prime }\nabla _{j}^{\prime }+\sum_{i}\sum_{j>1}\sum_{k>1}\frac{%
t_{ji}t_{ki}}{m_{i}}\nabla _{j}^{\prime }\nabla _{k}^{\prime }
\end{equation}
It is our purpose to keep the transformation (\ref{eq:r'->r}) as general as
possible so that it applies to all the nonadiabatic approaches discussed in
this review. In order to uncouple one of the new coordinates $\mathbf{r}%
_{1}^{\prime }$ from the remaining ones we require that the coefficients of
the transformation (\ref{eq:r'->r}) satisfy
\begin{equation}
\sum_{i}\frac{t_{ji}t_{1i}}{m_{i}}=0,\;j>1  \label{eq:t_CM_1}
\end{equation}
The new coordinates transform under translations as follows:
\begin{equation}
\hat{U}(\mathbf{a})\mathbf{r}_{j}^{\prime }\hat{U}(\mathbf{a})^{\dagger }=%
\mathbf{r}_{j}^{\prime }+\mathbf{a}\sum_{i}t_{ji}  \label{eq:Ur'U+}
\end{equation}
If we require that $\mathbf{r}_{1}^{\prime }$ transforms exactly as the
original variables and that the remaining $\mathbf{r}_{j}^{\prime }$ are
translation invariant we have
\begin{equation}
\sum_{i}t_{ji}=\delta _{j1}  \label{eq:t_CM_2}
\end{equation}
If we choose $t_{1i}=\xi m_{i}$, where $\xi $ is an arbitrary real number,
then Eq.~(\ref{eq:t_CM_1}) becomes Eq.~(\ref{eq:t_CM_2}) for $j>1$. If we
then substitute $t_{1i}=\xi m_{i}$ into Eq.~(\ref{eq:t_CM_2}) with $j=1$ we
conclude that
\begin{equation}
t_{1i}=\frac{m_{i}}{M},\;M=\sum_{i}m_{i}  \label{eq:t1i}
\end{equation}
and $\mathbf{r}_{1}^{\prime }$ results to be the well known coordinate of
the center of mass of the molecule\cite{MM46}. The choice of the
coefficients of the transformation (\ref{eq:r'->r}) for the remaining
variables $\mathbf{r}_{j}^{\prime }$ $j>1$ is arbitrary as long as they
satisfy Eq.~(\ref{eq:t_CM_2}).

Finally, the total Hamiltonian operator reads

\begin{eqnarray}
\hat{H} &=&-\frac{\hbar ^{2}}{2M}\nabla _{1}^{\prime 2}+\hat{H}_{M}
\nonumber \\
\hat{H}_{M} &=&-\frac{\hbar ^{2}}{2}\sum_{j>1}\sum_{k>1}\left( \sum_{i}\frac{%
t_{ji}t_{ki}}{m_{i}}\right) \nabla _{j}^{\prime }\nabla _{k}^{\prime }+\frac{%
1}{4\pi \epsilon _{0}}\sum_{i=1}^{N-1}\sum_{j=i+1}^{N}\frac{q_{i}q_{j}}{%
r_{ij}}  \label{eq:H_M}
\end{eqnarray}
where $\hat{H}_{M}$ is the internal or molecular Hamiltonian operator. The
explicit form of the interparticle distances $r_{ij}$ in terms of the new
coordinates $r_{k}^{\prime }$ may be rather cumbersome in the general case.
We will consider it in the particular applications discussed in subsequent
sections.

For future reference it is convenient to define the center of mass and
relative kinetic energy operators
\begin{eqnarray}
\hat{T}_{CM} &=&-\frac{\hbar ^{2}}{2M}\nabla _{1}^{\prime 2}  \label{eq:TCM}
\\
\hat{T}_{rel} &=&-\frac{\hbar ^{2}}{2}\sum_{j>1}\sum_{k>1}\left( \sum_{i}%
\frac{t_{ji}t_{ki}}{m_{i}}\right) \nabla _{j}^{\prime }\nabla _{k}^{\prime }
\label{eq:Trel}
\end{eqnarray}
respectively, so that $\hat{T}=\hat{T}_{CM}+\hat{T}_{rel},$ $\hat{H}_{M}=%
\hat{T}_{rel}+V$, and $\hat{H}=\hat{T}_{CM}+\hat{H}_{M}$.

The inverse transformation $\mathbf{t}^{-1}$ exists and gives us the old
coordinates in terms of the new ones:
\begin{equation}
\mathbf{r}_{i}=\sum_{j}\left( \mathbf{t}^{-1}\right) _{ij}\mathbf{r}%
_{j}^{\prime }  \label{eq:r->r'}
\end{equation}
According to equations (\ref{eq:Ur'U+}) and (\ref{eq:t_CM_2}) we have $\hat{U%
}(\mathbf{a})\mathbf{r}_{i}\hat{U}(\mathbf{a})^{\dagger }=\left( \mathbf{t}%
^{-1}\right) _{i1}\mathbf{a}+\mathbf{r}_{i}$ from which we conclude that
\begin{equation}
\left( \mathbf{t}^{-1}\right) _{i1}=1,\;i=1,2,\ldots ,N  \label{eq:(1/t)i1}
\end{equation}
In order to understand the meaning of this result notice that the momentum
conjugate to $\mathbf{r}_{i}^{\prime }$ is given by the transformation
\begin{equation}
\mathbf{\hat{p}}_{i}^{\prime }=\sum_{j}\left( \mathbf{t}^{-1}\right) _{ji}%
\mathbf{\hat{p}}_{j}  \label{eq:p'->p}
\end{equation}
so that the linear momentum of the center of mass
\begin{equation}
\mathbf{\hat{p}}_{1}^{\prime }=\sum_{j}\mathbf{\hat{p}}_{j}  \label{eq:p'_1}
\end{equation}
is precisely the total linear momentum of the molecule. We also appreciate
that $T_{CM}=\mathbf{\hat{p}}_{1}^{\prime 2}/(2M)$ and that he inverse
transformation of the momenta is (see Eq.~(\ref{eq:Nabla->Nabla'}))
\begin{equation}
\mathbf{p}_{j}=\sum_{i}\mathbf{t}_{ij}\mathbf{p}_{i}^{\prime }
\label{eq:p->p'}
\end{equation}

Eq. (\ref{eq:(1/t)i1}) is also relevant to the behaviour of the dipole
moment in internal coordinates. We have
\begin{equation}
\mathbf{\mu }=\sum_{i}q_{i}\mathbf{r}_{i}  \label{eq:DM}
\end{equation}
in the laboratory--fixed frame, and
\begin{equation}
\hat{U}(\mathbf{a})\mathbf{\mu }\hat{U}(\mathbf{a})^{\dagger }=\mathbf{\mu +}%
q\mathbf{a,\;}q\mathbf{=}\sum_{i}q_{i}  \label{eq:mu(a)}
\end{equation}
clearly shows that $\mathbf{\mu }$ is invariant under a translation of the
origin for a neutral molecule $q=0$. In internal coordinates we have
\begin{equation}
\mathbf{\mu }=\sum_{j}q_{j}^{\prime }\mathbf{r}_{j}^{\prime
},\;q_{j}^{\prime }=\sum_{i}q_{i}\left( \mathbf{t}^{-1}\right) _{ij}
\end{equation}
and $\hat{U}(\mathbf{a})\mathbf{\mu }\hat{U}(\mathbf{a})^{\dagger }=\mathbf{%
\mu +}q_{1}^{\prime }\mathbf{a}$ that is consistent with Eq. (\ref{eq:mu(a)}%
) because $q_{1}^{\prime }=q$ by virtue of Eq. (\ref{eq:(1/t)i1}).

We have kept the transformation (\ref{eq:r'->r}) as general as possible in
order to have a suitable expression of the molecular Hamiltonian that
applies to all the nonadiabatic approaches discussed in this review. In
what follows we illustrate some particular ways of determining the remaining
transformation coefficients. More precisely, we will choose a point in the
molecule as the origin of the new coordinate axis and refer the positions of
$N-1$ particles to it (remember that three coordinates have been reserved
for the location of the center of mass).

As an example, consider an arbitrary set of particle labels $I$ and choose
\begin{equation}
\mathbf{r}_{j}^{\prime }=\mathbf{r}_{j}-\frac{1}{M_{I}}\sum_{i\in I}m_{i}%
\mathbf{r}_{i},\;j>1,\;M_{I}=\sum_{i\in I}m_{i}  \label{eq:change_var_MI}
\end{equation}
That is to say:
\begin{equation}
t_{ji}=\delta _{ji}-\frac{m_{i}}{M_{I}}\delta _{iI},\;j>1,\;\delta
_{iI}=\left\{
\begin{array}{l}
1\text{ if }i \in I \\
0\text{ otherwise}
\end{array}
\right.
\end{equation}
These coefficients $t_{ji}$ already satisfy Eq.~(\ref{eq:t_CM_2}). If $I$ is
the set of nuclear labels then we refer the positions $\mathbf{r}%
_{j}^{\prime }$, $j>1$ to the nuclear center of mass. If the set $I$
contains only one nuclear label we refer the positions of the remaining
particles to that particular nucleus. Different authors have already chosen
one or the other coordinate origin as shown in sections \ref{sec:DM_NBO} and
\ref{sec:Mol-Fixed}.

If we take into account that
\begin{equation}
\tau _{jk}=\sum_{i}\frac{t_{ji}t_{ki}}{m_{i}}=\frac{\delta _{jk}}{m_{j}}-%
\frac{1}{M_{I}}\left( \delta _{kI}+\delta _{jI}\right) +\frac{1}{M_{I}}
\end{equation}
then we realize that the transformation (\ref{eq:change_var_MI}) uncouples
the coordinates of the particles that belong to $I$ from the remaining ones
in the kinetic operator:
\begin{equation}
\tau _{jk}=\left\{
\begin{array}{l}
\frac{\delta _{jk}}{m_{j}}-\frac{1}{M_{I}}\text{ if }j,k\in I \\
0\text{ if }j\in I\text{ and }k\notin I\text{ or }j\notin I\text{ and }k\in I
\\
\frac{\delta _{jk}}{m_{j}}+\frac{1}{M_{I}}\text{ if }j,k\notin I
\end{array}
\right.
\end{equation}

The equations derived in this section are sufficiently general to cover all
the strategies commonly followed in the separation of the center of mass
prior to the solution of the Schr\"{o}dinger equation for atoms and
molecules. Other general expressions for the translation--free internal
coordinates have already been discussed by Sutcliffe and Woolley\cite
{S92,S93,S94,SW05a}. It is a most important topic in the nonadiabatic
quantum--mechanical approach to atoms and molecules.

\section{Symmetry}

\label{sec:symmetry} In this section we discuss the symmetry of a molecular
system that is determined by all the operators that commute with its
Hamiltonian. For example, the total Hamiltonian operator commutes with the
total linear momentum
\begin{equation}
\lbrack \hat{H},\mathbf{\hat{p}}]=0,\;\mathbf{\hat{p}}=\sum_{j=1}^{N}\mathbf{%
\hat{p}}_{j}=\mathbf{\hat{p}}_{1}^{\prime }  \label{eq:[H,p]}
\end{equation}
and also with the total angular momentum
\begin{equation}
\lbrack \hat{H},\mathbf{\hat{J}}]=0,\;\mathbf{\hat{J}}=\sum_{j=1}^{N}\mathbf{%
\hat{J}}_{j}  \label{eq:[H,J]}
\end{equation}
It follows from equations (\ref{eq:r->r'}) and (\ref{eq:p->p'}) that
\begin{equation}
\mathbf{\hat{J}}=\sum_{i}\mathbf{\hat{r}}_{i}\times \mathbf{\hat{p}}%
_{i}=\sum_{i}\mathbf{\hat{r}}_{i}^{\prime }\times \mathbf{\hat{p}}%
_{i}^{\prime }=\mathbf{\hat{r}}_{1}^{\prime }\times \mathbf{\hat{p}}%
_{1}^{\prime }+\mathbf{\hat{J}}^{\prime }  \label{eq:J->J'}
\end{equation}
where $\mathbf{\hat{J}}^{\prime }$ is the internal or molecular angular
momentum. It is clear that
\begin{equation}
\lbrack \hat{H}_{M},\mathbf{\hat{J}}^{\prime }]=0,\;\mathbf{\hat{J}}^{\prime
}=\sum_{i=2}^{N}\mathbf{\hat{r}}_{i}^{\prime }\times \mathbf{\hat{p}}%
_{i}^{\prime }  \label{eq:[H_M,J']}
\end{equation}

The Hamiltonian operator is invariant under permutation of identical
particles. We may formally write this symmetry as
\begin{equation}
\hat{\mathcal{P}}\hat{H}\hat{\mathcal{P}}^{-1}=\hat{H}\Rightarrow [\hat{H},%
\hat{\mathcal{P}}]=0  \label{eq:[H,perm]}
\end{equation}
where $\hat{\mathcal{P}}$ stands for any permutation of electrons or
identical nuclei. Since $\hat{T}_{CM}$ is invariant under permutation of
identical particles we conclude that
\begin{equation}
\hat{\mathcal{P}}\hat{H}\hat{_{M}\mathcal{P}}^{-1}=\hat{H}_{M}\Rightarrow [%
\hat{H}_{M},\hat{\mathcal{P}}]=0  \label{eq:[H_M,perm]}
\end{equation}

Besides, the molecular Hamiltonian operator is also invariant under
inversion
\begin{equation}
\hat{\Im}\hat{H}_{M}\hat{\Im}^{-1}\Rightarrow [\hat{H}_{M},\hat{\Im}]=0
\label{eq:[H,Inv]}
\end{equation}
where the inversion operator $\hat{\Im}$ produces the transformations $\hat{%
\Im}\mathbf{\hat{r}}_{j}\hat{\Im}^{-1}=-\mathbf{\hat{r}}_{j}$ and $\hat{\Im}%
\mathbf{\hat{p}}_{j}\hat{\Im}^{-1}=-\mathbf{\hat{p}}_{j}$. Also notice that $%
[\mathbf{\hat{J}}^{\prime }\mathbf{,}\hat{\Im}]=0$, $[\mathbf{\hat{J}}%
^{\prime }\mathbf{,}\hat{\mathcal{P}}]=0$ and $[\hat{\Im},\hat{\mathcal{P}}%
]=0$ so that we can choose a stationary state $\Psi $ to be simultaneous
eigenfunction of such set of commuting operators:
\begin{eqnarray}
\hat{H}_{M}\Psi &=&E\Psi ,  \nonumber \\
\hat{J}^{\prime 2}\Psi &=&\hbar ^{2}J^{\prime }(J^{\prime }+1)\Psi
,\;J^{\prime }=0,1,\ldots  \nonumber \\
\hat{J}_{z}^{\prime }\Psi &=&\hbar M_{J}^{\prime }\Psi ,\;M_{J}^{\prime
}=0,\pm 1,\ldots ,\pm J^{\prime }  \nonumber \\
\hat{\mathcal{P}}\Psi &=&\sigma \Psi  \nonumber \\
\hat{\Im}\Psi &=&\pm \Psi  \label{eq:Opers_Psi}
\end{eqnarray}
where $\sigma =1$ or $\sigma =-1$ for identical bosons or fermions,
respectively. A somewhat more rigorous discussion of the kinematics and
dynamics of molecules was provided, for example, by Wolley\cite{W91}. The
simplified discussion in this section is sufficient for present purposes.

\section{The Hellmann--Feynman theorem}

\label{sec:H-F} In this section we derive the Hellmann--Feynman theorem for
optimized variational wavefunctions because it is most important for the
accurate calculation of dipole moments\cite{SO96,SBH86}.

Consider a trial function $\Phi $ and the variational energy $W$ given by
\begin{equation}
W\left\langle \Phi \right| \left. \Phi \right\rangle =\left\langle \Phi
\right| \hat{H}\left| \Phi \right\rangle  \label{eq:W_var}
\end{equation}
An arbitrary variation $\delta \Phi $ leads to
\begin{equation}
\delta W\left\langle \Phi \right| \left. \Phi \right\rangle =\left\langle
\delta \Phi \right| \hat{H}-W\left| \Phi \right\rangle +\left\langle \delta
\Phi \right| \hat{H}-W\left| \Phi \right\rangle ^{*}  \label{eq:dW}
\end{equation}
were the asterisk denotes complex conjugation. The right--hand side of this
equation vanishes when $\delta W=0$; that is to say, when $W$ is stationary
with respect to the infinitesimal change $\delta \Phi $ in the trial
function. The variational or optimal trial function satisfies this condition.

If the Hamiltonian operator depends on a parameter $\lambda $, which may be,
for example, a mass, charge, force constant, the strength of a external
field, etc., then the variational function will also depend on it.
Therefore, differentiation of Eq.~(\ref{eq:W_var}) with respect to $\lambda $
leads to
\begin{equation}
\frac{\partial W}{\partial \lambda }\left\langle \Phi \right| \left. \Phi
\right\rangle =\left\langle \Phi \right| \frac{\partial \hat{H}}{\partial
\lambda }\left| \Phi \right\rangle +\left\langle \frac{\partial \Phi }{%
\partial \lambda }\right| \hat{H}-W\left| \Phi \right\rangle +\left\langle
\frac{\partial \Phi }{\partial \lambda }\right| \hat{H}-W\left| \Phi
\right\rangle ^{*}  \label{eq:dw/dlam}
\end{equation}
It is clear from this equation that if the set of variations $\delta \Phi $
includes $\left( \partial \Phi /\partial \lambda \right) \delta \lambda $
then the optimized trial function $\Phi $ will satisfy the well known
Hellmann--Feynman theorem\cite{HM67,FC87}:
\begin{equation}
\frac{\partial W}{\partial \lambda }=\left\langle \frac{\partial \hat{H}}{%
\partial \lambda }\right\rangle  \label{eq:H-F}
\end{equation}
where $\left\langle \hat{A}\right\rangle =\left\langle \Phi |\hat{A}|\Phi
\right\rangle /\left\langle \Phi |\Phi \right\rangle $.

\section{The Born--Oppenheimer Approximation}

\label{sec:BO} Since the vast majority of quantum--mechanical studies of
molecular properties are based on the Born--Oppenheimer (BO) approximation%
\cite{BH54} we discuss it here with some detail, in spite of the fact that
we are more interested in non BO approaches. Before proceeding we mention
the curious fact that the modern and most useful form of that approach
appears to have been first developed by Slater\cite{S27} in his study of the
Helium atom as early as 1927. However, we will adopt the common practice to
attribute it to Born and Oppenheimer. In this section we begin with the
traditional procedure proposed by Slater\cite{S27} and Born and Huang\cite
{BH54} and then mention other authors' approaches and criticisms.

For simplicity and clarity we write the Schr\"{o}dinger equation for a
molecule as
\begin{equation}
\hat{H}\Psi _{\alpha }(\mathbf{r}^{e},\mathbf{r}^{n})=E_{\alpha }\Psi
_{\alpha }(\mathbf{r}^{e},\mathbf{r}^{n})  \label{eq:Schrodinger}
\end{equation}
where $\mathbf{r}^{e}$ and $\mathbf{r}^{n}$ denote all the electronic and
nuclear coordinates, respectively, and the subscript $\alpha $ stands for
the collection of all the quantum numbers necessary for the description of
the molecular state. We use the symbols $\hat{T}_{e}$, $\hat{T}_{n}$, $%
V_{ee} $,$V_{ne}$, $V_{nn}$ to indicate the kinetic energy operators for the
electrons and nuclei, and the electrons--electrons, nuclei--electrons, and
nuclei--nuclei Coulomb interactions, respectively:
\begin{equation}
\hat{H}=\hat{T}_{n}+\hat{H}_{e}+V_{nn},\;\hat{H}_{e}=\hat{T}%
_{e}+V_{ee}+V_{ne}  \label{eq:H_molec2}
\end{equation}
The BO approximation is based on the ansatz
\begin{equation}
\Psi _{\alpha }(\mathbf{r}^{e},\mathbf{r}^{n})=\sum_{j}\chi _{\alpha j}(%
\mathbf{r}^{n})\varphi _{j}(\mathbf{r}^{e},\mathbf{r}^{n})  \label{eq:Psi_BO}
\end{equation}
and we assume that $\left\langle \varphi _{j}\right. \left| \varphi
_{k}\right\rangle _{e}=\delta _{jk}$, where the subscript $e$ indicates
integration over electronic coordinates. It follows from equations (\ref
{eq:Schrodinger}), (\ref{eq:H_molec2}) and (\ref{eq:Psi_BO}) that
\begin{equation}
\left\langle \varphi _{k}\right| \hat{H}\left| \Psi _{\alpha }\right\rangle
_{e}=\hat{T}_{n}\chi _{\alpha k}+\sum_{j}\left( U_{kj}+W_{kj}\right) \chi
_{\alpha j}=E_{\alpha }\chi _{\alpha k}  \label{eq:BO_1}
\end{equation}
where
\begin{equation}
U_{kj}=\left\langle \varphi _{k}\right| \hat{H}_{e}\left| \varphi
_{j}\right\rangle _{e}+V_{nn}\delta _{kj},\;W_{kj}=\left\langle \varphi
_{k}\right| \left. [\hat{T}_{n},\varphi _{j}]\right\rangle _{e}
\end{equation}
We have introduced the formal commutator $[\hat{T}_{n},\varphi _{j}]$ with
the sole purpose of making the resulting equation more compact.

In the standard BO approximation we choose the functions $\varphi _{j}$ to
be eigenfunctions of the electronic Hamiltonian $\hat{H}_{e}$
\begin{equation}
\hat{H}_{e}\varphi _{j}(\mathbf{r}^{e},\mathbf{r}^{n})=E_{ej}(\mathbf{r}%
^{n})\varphi _{j}(\mathbf{r}^{e},\mathbf{r}^{n})  \label{eq:BO_elect}
\end{equation}
so that
\begin{equation}
U_{kj}(\mathbf{r}^{n})=U_{j}(\mathbf{r}^{n})\delta _{kj},\;U_{j}(\mathbf{r}%
^{n})=E_{ej}(\mathbf{r}^{n})+V_{nn}(\mathbf{r}^{n})
\end{equation}
In the first approximation we assume the couplings $W_{kj}$ between
``electronic states'' to be small and keep just one term for the ground
state
\begin{equation}
\Psi _{\alpha }(\mathbf{r}^{e},\mathbf{r}^{n})\approx \chi _{\alpha 0}(%
\mathbf{r}^{n})\varphi _{0}(\mathbf{r}^{e},\mathbf{r}^{n})
\label{eq:Psi_BO_app}
\end{equation}
and
\begin{equation}
\left[ \hat{T}_{n}+U_{0}(\mathbf{r}^{n})\right] \chi _{\alpha 0}(\mathbf{r}%
^{n})=E_{n}^{BO}\chi _{\alpha 0}(\mathbf{r}^{n})  \label{eq:BO_nuclear}
\end{equation}
The addition of the diagonal term $W_{00}$ to the nuclear operator in this
equation gives rise to the adiabatic approximation.

Sutcliffe and Wolley have criticized the main assumptions of this approach
in several papers\cite{WS77,W91,S93,S94,SW05a}. For example, they pointed
out that the functions in the expansion in the right--hand side of Eq. (\ref
{eq:Psi_BO}) are assumed to be square integrable, whereas the left--hand
side is not because we have not removed the motion of the center of mass.
Such criticism, which also applies to the approximate and widely used ansatz
(\ref{eq:Psi_BO_app}), does not appear to be justified because the nuclear
functions $\chi _{\alpha j}$ are not square integrable. According to those
authors another drawback of the BO approximation is that it does not take
into account the permutational symmetry of identical nuclei that are treated
as distinguishable particles clamped in space in order to define a framework
geometry unambiguously. However, in principle we can introduce such
permutational symmetry in the approximate wavefunction (\ref{eq:Psi_BO_app})
by means of appropriate projection operators like those mentioned below in
Sec.\-~\ref{sec:DM_NBO}. In our opinion the Born--Oppenheimer approximation
is a consistent way of deriving a suitable approximate solution to the
Schr\"{o}dinger equation.

It is interesting how some popular good books on quantum chemistry introduce
this subject. For example, Szabo and Ostlung\cite{SO96} state that ``Our
discussion of this approximation is qualitative. The quantitative aspects of
this approximation are clearly discussed by Sutcliffe'' and they give a
reference and simply write the single product (\ref{eq:Psi_BO_app}). On the
other hand, Pilar\cite{P68}, referring to a Hamiltonian operator like (\ref
{eq:H=T+V_full}), says: ``where it is assumed that all nuclear and
electronic coordinates have been referred to the center of mass of the
system''. However, that Hamiltonian operator exhibits all the electronic and
nuclear coordinates and no coupling terms coming from the separation of the
center of mass. Therefore, it cannot be the internal Hamiltonian operator as
we have already seen above. These are just two examples of how most authors
blindly accept the clamped nuclei approximation without further analysis.

In some cases, mostly for diatomic molecules, the separation of the motion
of the center of mass has been carried out rigorously prior to the
application of the BO approximation\cite{HM67}. Sutcliffe and Woolley have
already discussed this issue in detail in several papers\cite
{WS77,W91,S93,S94,SW05a}.

Although it is not customary to remove the motion of the center of mass of
the whole molecule before the application of the BO approximation, it
appears to be common practice to separate the nuclear center of mass at the
second stage (\ref{eq:BO_nuclear}). Since the nuclei are much heavier than
the electrons we may assume that the error is small. A rigorous discussion
of this issue as well as a simple exactly solvable example are given
elsewhere\cite{K97,F08}.

A most important by--product of the clamped--nuclei approximation is that it
enables one to introduce the familiar (classical) chemical concept of
molecular structure into the quantum--mechanical calculations. The molecular
geometry is given by the equilibrium nuclear configuration $\mathbf{r}%
_{eq}^{n}$ at the minimum of $U(\mathbf{r}^{n})$:
\begin{equation}
\left. \nabla U(\mathbf{r}^{n})\right| _{\mathbf{r}^{n}=\mathbf{r}%
_{eq}^{n}}=0  \label{eq:BO_eq}
\end{equation}
At this point it is worth noticing that $U(\mathbf{r}^{n})$ would not appear
in a straightforward rigorous solution of the Schr\"{o}dinger equation (\ref
{eq:Schrodinger}). Sutcliffe and Woolley\cite
{WS77,W91,S92,S93,S94,SW05b,SW05a} have also pointed out to the difficulty
of discussing molecular structure without the BO approximation (or any of
its variants).

\section{Interaction between a molecule and an external field}

\label{sec:Ext_field} In order to make the present review sufficiently
self--contained we briefly develop the main equations for the interaction
between a molecule and an external potential $\Phi (\mathbf{r})$. The energy
of that interaction is given by\cite{J90}
\begin{equation}
W=\int \rho (\mathbf{r)}\Phi (\mathbf{r})\,d\mathbf{r}
\end{equation}
where $\rho (\mathbf{r})$ is the molecular charge density. If the potential
varies slowly in the region where the charge density is nonzero we can
expand it in a Taylor series about the coordinate origin located somewhere
in the molecule:
\begin{equation}
\Phi (\mathbf{r})=\Phi (0)+\sum_{u}u\frac{\partial \Phi }{\partial u}(0)+%
\frac{1}{2}\sum_{u}\sum_{v}uv\frac{\partial ^{2}\Phi }{\partial u\partial v}%
(0)+\ldots
\end{equation}
where $u,v=x,y,z$. Since $\nabla ^{2}\Phi (\mathbf{r)=0}$ we subtract $%
r^{2}\nabla ^{2}\Phi (0\mathbf{)/}6$ from this equation and rewrite the
result as
\begin{equation}
\Phi (\mathbf{r})=\Phi (0)+\sum_{u}u\frac{\partial \Phi }{\partial u}(0)+%
\frac{1}{6}\sum_{u}\sum_{v}\left( 3uv-r^{2}\delta _{uv}\right) \frac{%
\partial ^{2}\Phi }{\partial u\partial v}(0)+\ldots
\end{equation}
If we now take into account that the external field is given by $\mathbf{F}(%
\mathbf{r})=-\nabla \Phi (\mathbf{r})$ we have
\begin{equation}
\Phi (\mathbf{r})=\Phi (0)-\mathbf{r\cdot F}(0)-\frac{1}{6}%
\sum_{u}\sum_{v}\left( 3uv-r^{2}\delta _{uv}\right) \frac{\partial F_{u}}{%
\partial v}(0)+\ldots
\end{equation}
For a set of point charges $q_{i}$ located at $\mathbf{r}_{i}$, $%
i=1,2,\ldots ,N$ we have
\begin{equation}
\rho (\mathbf{r})=\sum_{i=1}^{N}q_{i}\delta (\mathbf{r}-\mathbf{r}_{i})
\end{equation}
and the interaction energy becomes
\begin{equation}
W=q\Phi (0)-\mathbf{\mu }\cdot \mathbf{F}(0)-\frac{1}{6}\sum_{u}%
\sum_{v}Q_{uv}\frac{\partial F_{u}}{\partial v}(0)+\ldots
\label{eq:W(F)_classic}
\end{equation}
where the net charge $q$, the dipole moment $\mathbf{\mu }$, and the
quadrupole moment $\mathbf{Q}$ are given by
\begin{eqnarray}
q &=&\sum_{i=1}^{N}q_{i}  \nonumber \\
\mathbf{\mu } &=&\sum_{i=1}^{N}q_{i}\mathbf{r}_{i}  \nonumber \\
Q_{uv} &=&\sum_{i=1}^{N}q_{i}\left( 3u_{i}v_{i}-r_{i}^{2}\delta _{uv}\right)
\label{eq:multipoles}
\end{eqnarray}
If the applied field is uniform in the region of interest, then all the
terms beyond the first two ones in the right--hand side of Eq.~(\ref
{eq:W(F)_classic}) vanish.

In one of his well known discussions of long--range intermolecular forces
Buckingham\cite{B67} considers a Hamiltonian operator of the form
\begin{equation}
\hat{H}=\hat{H}^{0}-\mathbf{\mu }\cdot \mathbf{F}-\frac{1}{3}%
\sum_{u}\sum_{v}\Theta _{uv}F_{uv}+\ldots  \label{eq:H(F)}
\end{equation}
where $\mathbf{\Theta }=\mathbf{Q}/2$ and $F_{uv}=(\partial F_{u}/\partial
v)(0)$. Thus, the perturbation expansion for the energy of the molecule in
the external field results to be\cite{B67}
\begin{equation}
W=\left\langle \Psi \right| \hat{H}\left| \Psi \right\rangle
=W^{(0)}-\sum_{u}\mu _{u}^{(0)}F_{u}-\frac{1}{2}\sum_{u}\sum_{v}\alpha
_{uv}F_{u}F_{v}-\ldots -\frac{1}{3}\sum_{u}\sum_{v}\Theta
_{uv}^{(0)}F_{uv}-\ldots  \label{eq:W(F)_expansion}
\end{equation}
where
\begin{eqnarray}
\mu _{u}^{(0)} &=&\left\langle \Psi ^{(0)}\right| \hat{\mu}_{u}\left| \Psi
^{(0)}\right\rangle  \nonumber \\
\Theta _{uv}^{(0)} &=&\left\langle \Psi ^{(0)}\right| \hat{\Theta}%
_{uv}\left| \Psi ^{(0)}\right\rangle  \label{eq:<mu>}
\end{eqnarray}
are the permanent dipole and quadrupole moments, respectively. Buckingham%
\cite{B67} does not write the Hamiltonian operator $\hat{H}^{(0)}$
explicitly, but he refers to the energy of ``separate molecules for fixed
molecular positions and orientations''. Therefore one may assume that he
probably means the BO Hamiltonian operator. On the other hand, Bishop\cite
{B90} considers that Eq.~(\ref{eq:W(F)_expansion}) can be used irrespective
of whether one is considering the electronic or the total molecular energy.
He explicitly indicates that $q_{i}$ is an element of charge at the point $%
\mathbf{r}_{i}$ relative to an origin fixed at some point in the molecule.
As shown below in sections \ref{sec:DM_NBO} and \ref{sec:Mol-Fixed}, this
choice of reference frame is not convenient for the nonadiabatic calculation
of the dipole moment.

In a most interesting and comprehensive review of the electric moments of
molecules Buckingham\cite{B70} argues, by means of simple and rigorous
symmetry arguments, that a diatomic molecule in some stationary states does
not possess a dipole moment. We will come back to this point in sections~\ref
{sec:DM_NBO} and \ref{sec:Mol-Fixed}.

\section{Experimental Measurements of Dipole Moments}

\label{sec:exp} Typical experimental measurements of dipole moments are
based on microwave spectra, and almost invariably on the Stark effect and
the model of a rigid (or almost rigid) rotating dipole. If $\hat{H}_{rot}$
is the hamiltonian of a rigid rotator and $-\mathbf{\mu }\cdot \mathbf{F}$
is the interaction between the molecular dipole moment and the uniform
electric field (as shown above in Sec.~\ref{sec:Ext_field})), then the Stark
rotational energies are given by
\begin{equation}
\hat{H}_{rot}(F)\psi ^{rot}=E^{rot}(F)\psi ^{rot},\;\hat{H}_{rot}(F)=\hat{H}%
_{rot}(0)-\mu F\cos (\theta )  \label{eq:rot_energy}
\end{equation}
where $\theta $ is the angle between the external field $\mathbf{F}$ and the
dipole moment $\mathbf{\mu }$.

The external electric field produces both shift and splitting of the
molecular rotational energies.\ The magnitude of the shift of the spectral
lines $\tilde{\nu}(F)=\Delta E^{rot}(F)/(hc)$ changes with the field.
Rayleigh--Schr\"{o}dinger perturbation theory provides analytical
expressions for the Stark shifts in the form of a power series of the field
intensity:
\begin{equation}
\Delta \tilde{\nu}(F)=\tilde{\nu}(F)-\tilde{\nu}(0)=\tilde{\nu}^{(1)}\mu F+%
\tilde{\nu}^{(2)}(\mu F)^{2}+\ldots  \label{eq:Stark_shift}
\end{equation}
where the coefficients $\tilde{\nu}^{(j)}$ are known functions of the
rotational quantum numbers\cite{R56,GST53,TS55}. If we measure the Stark
shifts for known values of the field intensity and then fit selected
experimental data to a polynomial function of the field we can obtain the
dipole moment from the polynomial coefficients.

In order to account for the hyperfine structure of the spectra one should
add the terms arising from the spin--rotational interaction, the quadrupole
interaction of the nuclei, and the spin--spin magnetic interactions\cite
{R56,GST53,TS55}.

Molecular beam electric resonance experiments are also based on the same
model of rotating dipole\cite{R56}.

There have been many experimental studies of a wide variety of molecules. We
restrict to the simplest ones that are accessible to existing nonadiabatic
approaches, such as, obviously, diatomics\cite{WGK60,WGK62,R69,MK70}. In
this case $\tilde{\nu}^{(1)}=0$ and the Stark effect is quadratic. It is
worth paying attention to the discrepancy in notation and presentation of
the empirical models. For example, the model Hamiltonian may be either an
actual operator\cite{R69} or a scalar function of the quantum numbers\cite
{MK70}. The interaction between the dipole and the field may also be written
in somewhat different ways\cite{R69,MK70} or the effective Hamiltonian may
omit the rotational kinetic energy\cite{WGK62}.

Although symmetric tops appear to be beyond present nonadiabatic treatments,
we quote them here as another example of the use of the model of a rotating
dipole outlined above\cite{LG63,SG66}. In this case $\tilde{\nu}^{(1)}=0$
when $K=0$ and $\tilde{\nu}^{(1)}\neq 0$ when $K\neq 0$, where $K$ is the
quantum number for the projection of the angular momentum along the symmetry
axis.

It is clear from the discussion above that the accuracy and reliability of
the experimental determination of the molecular dipole moment does not
depend only on the accuracy of the measured Stark line shifts but also on
the validity of the theoretical, semi--empirical model of the molecule as a
rotating electric dipole. We will come back on this important point later in
Sec.~\ref{sec:PT}.

\section{Born--Oppenheimer Calculations of Dipole Moments}

\label{sec:DM_BO} Although we are mainly interested in the nonadiabatic
calculation of dipole moments it is worth comparing it with the more popular
BO approach to the problem. For this reason, in this section we outline the
latter. If we are able to solve Eq. (\ref{eq:BO_elect}) for an appropriate
set of nuclear configurations and determine the equilibrium geometry of the
molecule given by Eq. (\ref{eq:BO_eq}), we can then calculate the dipole
moment as follows:
\begin{equation}
\mathbf{\mu }^{BO}=-e\int \varphi (\mathbf{r}^{e},\mathbf{r}%
_{eq}^{n})^{*}\left( \sum_{j=1}^{N_{e}}\mathbf{r}_{j}^{e}\right) \varphi (%
\mathbf{r}^{e},\mathbf{r}_{eq}^{n})\,d\mathbf{r}^{e}+e\sum_{j=1}^{N_{n}}Z_{j}%
\mathbf{r}_{j\,eq}^{n}  \label{eq:BO_DM}
\end{equation}
for a molecule with $N_{e}$ electrons and $N_{n}$ nuclei\cite{SO96}. In this
equation $d\mathbf{r}^{e}$ denotes the volume element for all the electronic
coordinates.

An alternative approach consists in solving the electronic BO equation for
the molecule in an electric field
\begin{equation}
\hat{H}_{e}(\mathbf{F})=\hat{H}_{e}(0)+e\mathbf{F}\cdot \sum_{j=1}^{N_{e}}%
\mathbf{r}_{j}^{e}  \label{eq:BO_H_e(F)}
\end{equation}
and then calculate the electronic part of the dipole moment as
\begin{equation}
\mu _{u}^{e}=-\left( \frac{\partial E_{e}(\mathbf{r}_{eq}^{n},\mathbf{F})}{%
\partial F_{u}}\right) _{F=0}  \label{eq:BO_DM_e}
\end{equation}
where $E_{e}(\mathbf{r}_{eq}^{n},\mathbf{F})$ is the lowest eigenvalue of
the electronic operator (\ref{eq:BO_H_e(F)}) for the equilibrium geometry.

Szabo and Ostlund\cite{SO96} discuss the reasons for the noticeable
discrepancy between both approaches. It is known to be due to the fact that
approximate wavefunctions which are not well optimized fail to satisfy the
Hellmann-Feynman theorem\cite{SBH86} already discussed in Sec.~\ref{sec:H-F}%
. Since the formally correct definition of properties like the dipole moment
is as a response function to an external field (see Sec.~\ref{sec:Ext_field}%
) Swanton et al\cite{SBH86} proposed a calculation based on the following
expression
\begin{equation}
\mu _{u}^{e}=\left\langle \Phi (0)\right| \hat{\mu}_{u}^{e}\left| \Phi
(0)\right\rangle -2\left\langle \frac{\partial \Phi }{\partial F_{u}}%
(0)\right| [\hat{H}_{e}(0)-E_{e}(\mathbf{r}_{eq}^{n},0)\left| \Phi
(0)\right\rangle  \label{eq:BO_DM_e_2}
\end{equation}
in such cases where the variational function is not fully optimized and one
does not have appropriate analytical expressions for the derivatives $%
(\partial E_{e}/\partial F_{u})_{F=0}$. This equation is a particular case
of (\ref{eq:dw/dlam}) when the electronic approximate wavefunction is real
and normalized to unity $\left\langle \Phi (0)\right| \left. \Phi
(0)\right\rangle =1$.

The nuclear configuration given by the set of equilibrium coordinates $%
\mathbf{r}_{eq}^{n}$ determines what we usually call the geometry of the
molecule and thereby the orientation of the dipole moment. For example, we
know that the dipole moment is directed along the molecular axis in a linear
molecule without an inversion center or along the symmetry axis in a
symmetric top. The analysis is not so simple in the case of the nonadiabatic
calculations that we will discuss in Sec.~\ref{sec:DM_NBO}.

According to Cade and Huo\cite{CH66} the dipole moment calculated by means
of this quantum--mechanical approach for just one internuclear distance
(say, the theoretical or experimental $R_{e}$) is not strictly comparable to
the experimental one that commonly corresponds to a particular vibrational
state or an average over a set of vibrational states. In spite of this
apparent deficiency of the BO approach, it is worth noticing that when
Wharton et al\cite{WGK60} determined the dipole moment of LiH experimentally
to be $\mu =5.9D$ there were as many as eight previous reliable
quantum--mechanical calculations that agreed with it to $\pm 0.3D$\cite{M60}.

\section{Nonadiabatic Calculations of Dipole Moments}

\label{sec:DM_NBO} Most theoretical calculations of dipole moments are based
on the BO approximation for several reasons: first, and most importantly,
non BO calculations require far more computer time, second, most theoretical
chemists are unwilling to go beyond the BO approximation that they assume to
know well, third, non BO calculations give rise to some additional
theoretical difficulties as we will show below. It is therefore not
surprising that non BO calculations of dipole moments are scarce and few. We
describe some of then in this section and in Sec.~\ref{sec:Mol-Fixed}.

Tachikawa and Osamura\cite{TO00} proposed a dynamic extended molecular
orbital (DEMO) approach based on SCF wavefunctions of the form
\begin{equation}
\Psi ^{SCF}=\prod_{I}\Phi ^{I}  \label{eq:NBO_SCF}
\end{equation}
where each $\Phi ^{I}$ is a function of the coordinates of a set of
identical particles with the appropriate permutation symmetry. These
functions are expressed in terms of generalized molecular orbitals $\phi
_{j}^{I}$ that are linear combinations of floating Gaussians $\chi _{r}^{I}$%
:
\begin{equation}
\phi _{j}^{I}=\sum_{r}c_{rj}^{I}\chi _{r}^{I}  \label{eq:DEMO_FG}
\end{equation}
It is worth noticing that Tachikawa and Osamura[TO00] did not separate the
motion of the center of mass, and this omission will give rise to
considerable errors if the SCF orbitals depend on laboratory--fixed
coordinates\cite{F09b}. Another limitation of this approach is that the SCF
wavefunction does not take into account particle correlation that may be
quite strong between nuclei\cite{CA02a,CADL03}.

If the variational approach were based on a trial function $\varphi (\mathbf{%
r}_{2}^{\prime },\mathbf{r}_{3}^{\prime },\ldots ,\mathbf{r}_{N}^{\prime })$
of just internal, translation--free, coordinates $\mathbf{r}_{j}^{\prime }$
(see Sec.~\ref{sec:Hamiltonian}), it would not be necessary to separate the
motion of the center of mass explicitly because $\left\langle \varphi
\right| \hat{H}\left| \varphi \right\rangle =\left\langle \varphi \right|
\hat{H}_{M}\left| \varphi \right\rangle $. But it is not the case of the SCF
wavefunction (\ref{eq:NBO_SCF}) so that we have $E^{SCF}=\left\langle \Psi
^{SCF}\right| \hat{H}\left| \Psi ^{SCF}\right\rangle =\left\langle \Psi
^{SCF}\right| \hat{T}_{CM}\left| \Psi ^{SCF}\right\rangle +\left\langle \Psi
^{SCF}\right| \hat{H}_{M}\left| \Psi ^{SCF}\right\rangle >$ $\left\langle
\Psi ^{SCF}\right| \hat{H}_{M}\left| \Psi ^{SCF}\right\rangle $. Therefore,
the estimated energy will always be worse than when the trial function
depends only on internal coordinates, even though the SCF wavefunction may
satisfy the virial theorem $2\left\langle \Psi ^{SCF}\right| \hat{T}\left|
\Psi ^{SCF}\right\rangle =-\left\langle \Psi ^{SCF}\right| V\left| \Psi
^{SCF}\right\rangle $\cite{FC87,TO00}. The reader may find a more detailed
discussion of this issue elsewhere\cite{F09b}.

Tachikawa and Osamura\cite{TO00} calculated the dipole moments of the $^{m}$H%
$^{n}$H and $^{m}$Li$^{n}$H isotopomer series, but, unfortunately, they did
not show the expression that they used. This issue is not a minor one as
discussed in what follows.

Before proceeding further, it is convenient to discuss the failure of the
naive approach to the nonadiabatic calculation of dipole moments. In quantum
mechanics one obtains the average of an observable $O$ as the expectation
value $\left\langle \Psi \right| \hat{O}\left| \Psi \right\rangle $ of the
corresponding operator $\hat{O}$. In Sec.~\ref{sec:symmetry} we showed that $%
\hat{\Im}\Psi =\pm \Psi $ because the molecular Hamiltonian is invariant
under inversion. Since $\hat{\Im}\mathbf{\hat{\mu}}\hat{\Im}^{-1}=-\mathbf{%
\hat{\mu}}$ we conclude that
\begin{equation}
\mathbf{\mu }=\left\langle \Psi \right| \mathbf{\hat{\mu}}\left| \Psi
\right\rangle =0  \label{eq:<mu>=0}
\end{equation}
for any nondegenerate molecular state $\Psi $. This result that is known
since long ago\cite{B70} applies to \textit{any} molecule in its ground
state and renders fruitless the calculation of its dipole moment as a
straightforward expectation value in a set of axes parallel to the
laboratory one (like the one discussed in Sec.~\ref{sec:Hamiltonian}).

One of the problems that arises from the separation of the center of mass
discussed in Sec.~\ref{sec:Hamiltonian} is that the transformation (\ref
{eq:r->r'}) can make the Coulomb potential rather messy. One way of keeping
a simple form of the potential--energy function is to choose one of the
particles as coordinate origin. From a practical point of view it appears to
be convenient to choose the heaviest nucleus for that purpose\cite{CA02a}.
Thus, the transformation
\begin{eqnarray}
\mathbf{r}_{1}^{\prime } &=&\sum_{i=1}^{N}\frac{m_{i}}{M}\mathbf{r}_{i}
\nonumber \\
\mathbf{r}_{j}^{\prime } &=&\mathbf{r}_{j}-\mathbf{r}_{1},\;j=2,3,\ldots ,N
\label{eq:change_var_r1}
\end{eqnarray}
leads to
\begin{equation}
\hat{H}_{M}=-\frac{\hbar ^{2}}{2}\sum_{i}\frac{1}{m_{i}}\nabla _{i}^{\prime
2}-\frac{\hbar ^{2}}{2m_{1}}\sum_{j>1}\sum_{k>1}\nabla _{j}^{\prime }\nabla
_{k}^{\prime }+\frac{1}{4\pi \epsilon _{0}}\sum_{i=1}^{N-1}\sum_{j=i+1}^{N}%
\frac{q_{i}q_{j}}{r_{ij}}  \label{eq:H_M_2}
\end{equation}
where $m_{1}$ is the mass of the heaviest nucleus and $\mathbf{r}_{1}$ its
location in the laboratory reference frame. The transformed Coulomb
potential does not make the calculation of matrix elements unnecessarily
complicated as follows from the fact that $|\mathbf{r}_{i}-\mathbf{r}_{1}|=|%
\mathbf{r}_{i}^{\prime }|$, $i=2,3,\ldots ,N$ and $|\mathbf{r}_{i}-\mathbf{r}%
_{j}|=|\mathbf{r}_{i}^{\prime }-\mathbf{r}_{j}^{\prime }|$, $i,j=2,3,\ldots
,N$.

As an example we consider a four--particle molecule. The transformation (\ref
{eq:change_var_r1}) is given by the matrix
\begin{equation}
\mathbf{t}=\left(
\begin{array}{cccc}
\frac{m_{1}}{M} & \frac{m_{2}}{M} & \frac{m_{3}}{M} & \frac{m_{4}}{M} \\
-1 & 1 & 0 & 0 \\
-1 & 0 & 1 & 0 \\
-1 & 0 & 0 & 1
\end{array}
\right)  \label{eq:t(4)}
\end{equation}
with inverse
\begin{equation}
\mathbf{t}^{-1}=\left(
\begin{array}{cccc}
1 & -\frac{m_{2}}{M} & -\frac{m_{3}}{M} & -\frac{m_{4}}{M} \\
1 & 1-\frac{m_{2}}{M} & -\frac{m_{3}}{M} & -\frac{m_{4}}{M} \\
1 & -\frac{m_{2}}{M} & 1-\frac{m_{3}}{M} & -\frac{m_{4}}{M} \\
1 & -\frac{m_{2}}{M} & -\frac{m_{3}}{M} & 1-\frac{m_{4}}{M}
\end{array}
\right)  \label{eq:1/t(4)}
\end{equation}
This transformation applies, for example, to the H$_{2}$ isotopomer series%
\cite{CA02c}. The only members that exhibit dipole moments are HD, HT and
DT. In this case we choose the labels 1,2,3,4 to denote the heaviest
nucleus, the lightest one, and the two electrons, respectively.

The fact that $m_{1}>m_{2}$ gives rise to a charge asymmetry and a small
dipole moment. Obviously, the straightforward BO approximation cannot
account for it because the electronic Hamiltonian $\hat{H}_{e}$ does not
depend on the nuclear masses (see Sec.~\ref{sec:BO}) and the resulting
electronic charge density is exactly the same for all isotopomers.
Therefore, in order to obtain the dipole moment of diatomic molecules of the
form $^{m}$A$^{n}$A, one has to take into account nonadiabatic corrections%
\cite{B60,B61,FB77}. An alternative approach is based on the fact that there
is no unique way of implementing the BO approximation. In fact, appropriate
canonical transformations of the coordinates prior to the application of the
BO approximation may force the required asymmetry and provide a suitable way
of calculating the dipole moment of $^{m}$A$^{n}$A molecules\cite{TCK85a,
TCK85b}.

Cafiero and Adamowicz\cite{CA02a,CA02c} calculated nonadiabatic dipole
moments for some small diatomic molecules; in what follows we outline the
variational method proposed by those authors for a diatomic molecule with $%
N-2$ electrons. The core of the approach is a basis set of floating $s$%
--type explicitly correlated Gaussian functions of the form
\begin{eqnarray}
g_{k}(\mathbf{r}) &=&\exp \left[ -(\mathbf{X}-\mathbf{s}_{kx})\cdot \mathbf{A%
}_{k}\cdot (\mathbf{X}-\mathbf{s}_{kx})^{t}-(\mathbf{Y}-\mathbf{s}%
_{ky})\cdot \mathbf{A}_{k}\cdot (\mathbf{Y}-\mathbf{s}_{ky})^{t}\right.
\nonumber \\
&&\left. -(\mathbf{Z}-\mathbf{s}_{kz})\cdot \mathbf{A}_{k}\cdot (\mathbf{Z}-%
\mathbf{s}_{kz})^{t}\right]  \label{eq:gk(r)}
\end{eqnarray}
where $\mathbf{A}_{k}$ is a $N^{\prime }\times N^{\prime }$ ($N^{\prime
}=N-1)$ symmetric matrix, $\mathbf{X}$, $\mathbf{Y}$, and $\mathbf{Z}$ are $%
1\times N^{\prime }$ matrices of the form $\mathbf{X}=(x_{2}^{\prime
},x_{3}^{\prime },\ldots ,x_{N}^{\prime })$, $\mathbf{s}_{k}$ is a $1\times
N^{\prime }$ matrix that determines the location of the center of the
Gaussian in space and $t$ stands for transpose. In order to assure that the
Gaussians are square integrable they chose $\mathbf{A}_{k}$ to be of
Cholesky factored form $\mathbf{A}_{k}=\mathbf{L}_{k}\cdot \mathbf{L}%
_{k}^{t} $, where $\mathbf{L}_{k}$ is a $N^{\prime }\times N^{\prime }$
lower triangular matrix.

In order to have the correct permutation symmetry of identical particles
they resorted to appropriate projection operators of the form\cite{CA02a}
\begin{equation}
\hat{E}=\prod_{i}\hat{E}_{i}  \label{eq:E_proj}
\end{equation}
and constructed the variational function
\begin{equation}
\Psi =\sum_{k=1}^{m}c_{k}\hat{E}g_{k}(\mathbf{r})  \label{eq:Psi_CA}
\end{equation}
so that the permutation of any pair of identical particles leads to either $%
\Psi $ or $-\Psi $ if they are bosons or fermions, respectively. Then, they
minimized the variational energy
\begin{equation}
E=\min \frac{\left\langle \Psi \right| \hat{H}\left| \Psi \right\rangle }{%
\left\langle \Psi \right| \left. \Psi \right\rangle }  \label{eq:E_CA}
\end{equation}
Notice that the matrix $\mathbf{L}_{k}$ has $N^{\prime }(N^{\prime }+1)/2$
independent variational parameters and each $\mathbf{s}_{k}$ contributes
with $N^{\prime }$; therefore, there are $N^{\prime }(N^{\prime
}+1)/2+3N^{\prime }+1$ adjustable variational parameters for every basis
function. In a most comprehensive review Bubin et al\cite{BCA05} discussed
this type of calculation in detail.

Since the straightforward expectation value of the dipole--moment operator $%
\mathbf{\hat{\mu}}$ will not produce any physically meaningful result, as
discussed above in this section, Cafiero and Adamowicz\cite
{CA02a,CA02c,CADL03} resorted to an alternative approach based on the energy
of the molecule in an external electric field $\mathbf{F}$, given by the
Hamiltonian operator:
\begin{equation}
\hat{H}(\mathbf{F})=\hat{H}_{M}-\mathbf{F}\cdot \mathbf{\hat{\mu}}
\label{eq:H_CA}
\end{equation}
They fitted energy values to a polynomial function of the field\cite
{CA02a,CA02c,CADL03}:
\begin{equation}
E(F_{z})=E(0)-\mu _{z}F_{z}-\frac{1}{2}\alpha _{zz}F_{z}^{2}-\ldots
\label{eq:E(F)_poly}
\end{equation}
and obtained the dipole moment from the linear term. The dipole moments of
HD, HT, LiH and LiD calculated in this way proved to be very accurate and in
remarkable agreement with available experimental values\cite{CA02a,CA02c}.
In particular, the rate of convergence of the theoretical results for the LiH%
\cite{CA02a} towards the corresponding experimental dipole moment\cite{WGK62}
is astonishing. Table~\ref{tab:DM} shows some dipole moments calculated by
Cafiero and Adamowicz\cite{CA02a,CA02c} and other authors\cite
{PWHU96,TO00,KW66} as well as the corresponding experimental values\cite
{LT74,WGK62}.

However, if the approximate variational function (\ref{eq:Psi_CA}) were
fully optimized, then it should satisfy the Hellmann--Feynman theorem
discussed in Sec.~\ref{sec:H-F} and in that case the only possible result
would be
\begin{equation}
\mu _{z}=\left. \frac{\partial E(F_{z})}{\partial F_{z}}\right|
_{F_{z}=0}=\left\langle \Psi \right| \hat{\mu}_{z}\left| \Psi \right\rangle
=0  \label{eq:<mu>_HF}
\end{equation}
The only way that Cafiero and Adamowicz\cite{CA02a,CA02c} could obtain a
nonzero dipole moment by means of this approach is that their variational
wavefunction did not become spherically symmetric when $F_{z}\rightarrow 0$.
In other words, that their variational ansatz was not sufficiently accurate
at small values of the field strength where it would reveal the permanent
molecular dipole moment. In that case, it is not clear how those authors\cite
{CA02a,CA02c} obtained such remarkable agreement between their theoretical
results and the experimental values of the dipole moment\cite{WGK62}. A more
detailed discussion of this baffling agreement has been published elsewhere%
\cite{F08b,F09}. A most speculative explanation is that a biased placement
of the centers of the floating Gaussians somehow mimicked the use of a
body--fixed set of axes (see Sec.~\ref{sec:Mol-Fixed}). However, the authors
never mentioned this possibility. Curiously, in a later paper the authors
state that ``This spherical symmetry for the ground--state wavefunction
implies several things about molecules that may go against common chemical
intuition. First of all, no molecule in the ground state will have a dipole
moment, just as atoms do not. Similarly, the molecule will have only one
unique polarizability, an isotropic polarizability. The current authors have
presented several papers which discuss these phenomena''\cite{CA04}.
However, they did not explain how they had obtained the dipole moments in
two of those earlier papers\cite{CA02a,CA02c}. Besides, in Sec.~\ref
{sec:Mol-Fixed} we will show that the statement ``no molecule in the ground
state will have a dipole moment, just as atoms do not'' is false.

It would be interesting to investigate to which extent the variational
ansatz (\ref{eq:Psi_CA}) optimized by Cafiero and Adamowicz\cite{CA02a,CA02c}
satisfies the Hellmann--Feynman theorem that in the present case gives us an
exact theoretical relation between the molecule's response to the field and
the dipole moment. Exact relationships like those given by the hypervirial
and Hellmann--Feynman theorems are useful to determine the accuracy of
approximate wavefunctions\cite{HM67,FC87}. The questionable success of the
method devised by those authors is obviously based on the floating nature of
the Gaussian functions that one can place conveniently to get the desired
result\cite{CA02a,CA02c}. Notice that Bubin et al\cite{BLSA09} resorted to
one--center Gaussian functions of the form
\begin{equation}
\phi _{k}(\mathbf{r})=r_{1}^{m_{k}}\exp \left[ -\mathbf{X}\cdot \mathbf{A}%
_{k}\cdot \mathbf{X}^{t}-\mathbf{Y}\cdot \mathbf{A}_{k}\cdot \mathbf{Y}^{t}-%
\mathbf{Z}\cdot \mathbf{A}_{k}\cdot \mathbf{Z}^{t}\right]  \label{eq:gk(r)_2}
\end{equation}
to determine the charge asymmetry of the rotationless states of the HD
molecule. Clearly, this kind of basis functions cannot be placed at will to
force an axial symmetry and a nonzero dipole moment. It is for this reason
that we mentioned above in this section that it is unfortunate that
Tachikawa and Osamura\cite{TO00} did not show the expression that they used
for the calculation of the dipole moment of the $^{m}$H$^{n}$H and $^{m}$Li$%
^{n}$H molecules.

The nonadiabatic calculations of molecular dipole moments described in this
section are bound to fail because they are based on a set of axes parallel
to the laboratory one. Therefore, the expectation value of the
dipole--moment operator vanishes for any molecule in a nondegenerate state.
In Sec.~\ref{sec:Mol-Fixed} we discuss more judicious calculations based on
molecule--fixed coordinate systems.

\section{Molecule--Fixed Coordinate System}

\label{sec:Mol-Fixed} It is convenient to discuss the motion of a system of
particles in space by means of three sets of axes. One set of axis remains
fixed somewhere in the laboratory. A second set of axes with its origin
fixed on the system center of mass and parallel to the laboratory one. It
was discussed in Sec.~\ref{sec:Hamiltonian} with the purpose of separating
the motion of the entire system as a point particle with mass equal to the
total mass of the system. A third set of axes with origin on the center of
mass and somehow completely fastened to the system enables us to describe
its rotational motion. It is not difficult to define such set of axes for a
rigid body, but it is not so obvious when the distance between the system
particles change rather arbitrarily. It is clear that we should define any
intrinsic molecular property, like the electric dipole moment, moment of
inertia, etc, with respect to this body--fixed reference frame. If we use
the second set of axes we expect that the expectation value of the
dipole--moment operator for the molecule in a nondegenerate state vanishes
as discussed in Sec.~\ref{sec:DM_NBO}. This result is an obvious consequence
of the average over the angular degrees of freedom of the whole system. The
proper use of the body--fixed set of axes was clearly addressed by Blinder%
\cite{B60,B61} in his studies on the HD molecule, and Sutcliffe\cite
{S92,S93,S94,S99} discussed the more general case of polyatomic molecules.
For simplicity, we restrict ourselves to Blinder's proposal for diatomic
molecules\cite{B60,B61} in what follows.

For generality we first consider a diatomic molecule with the nuclei located
at $\mathbf{r}_{1}$ and $\mathbf{r}_{2}$ and the electrons at $\mathbf{r}%
_{2},\mathbf{r}_{3},\ldots ,\mathbf{r}_{N}$. Following Blinder\cite{B60} we
use relative coordinates for the nuclei and refer the electron coordinates
to the midpoint between the nuclei according to
\begin{eqnarray}
\mathbf{r}_{1}^{\prime } &=&\frac{1}{M}\sum_{i}m_{i}\mathbf{r}_{i}  \nonumber
\\
\mathbf{r}_{2}^{\prime } &=&\mathbf{r}_{2}-\mathbf{r}_{1}  \nonumber \\
\mathbf{r}_{i}^{\prime } &=&\mathbf{r}_{i}-\frac{1}{2}(\mathbf{r}_{1}+%
\mathbf{r}_{2}),\;i>2  \label{eq:transf_Blinder}
\end{eqnarray}
that is a particular case of Eq. (\ref{eq:r'->r}) with
\begin{eqnarray}
t_{1j} &=&\frac{m_{j}}{M}  \nonumber \\
t_{2j} &=&\delta _{j2}-\delta _{j1}  \nonumber \\
t_{ij} &=&\delta _{ij}-\frac{1}{2}(\delta _{j1}+\delta _{j2}),\;j>2
\label{eq:t_Blinder}
\end{eqnarray}
The resulting molecular Hamiltonian operator in the coordinate
representation will be
\begin{equation}
\hat{H}_{M}=-\frac{\hbar ^{2}}{2m_{s}}\nabla _{2}^{\prime 2}-\frac{\hbar ^{2}%
}{2m_{e}}\sum_{j=3}^{N}\nabla _{j}^{\prime 2}-\frac{\hbar ^{2}}{8m_{s}}%
\left( \sum_{j=3}^{N}\nabla _{j}^{\prime }\right) ^{2}+\frac{\hbar ^{2}}{%
2m_{a}}\nabla _{2}^{\prime }\cdot \sum_{j=3}^{N}\nabla _{j}^{\prime }+V
\label{eq:HM_diat}
\end{equation}
where $m_{e}$ is the electronic mass, $m_{s}=m_{1}m_{2}/(m_{1}+m_{2})$ and $%
m_{a}=m_{1}m_{2}/(m_{1}-m_{2})$. These coordinates are most convenient to
calculate the dipole moment of diatomic molecules of the form $^{m}$A$^{n}$A
because the fourth term in the right--hand side of the internal Hamiltonian (%
\ref{eq:HM_diat}) vanishes when $m_{1}=m_{2}$ and can therefore be treated
as a perturbation that converts the symmetric case to an asymmetric one.

In order to place the molecule--fixed set of axes we take into account the
unit vectors generated by
\begin{equation}
\mathbf{R}=\mathbf{r}_{2}-\mathbf{r}_{1}=R(\cos \phi \sin \theta \,\mathbf{e}%
_{x}+\sin \phi \sin \theta \,\mathbf{e}_{y}+\cos \theta \,\mathbf{e}_{z})
\label{eq:R_Blinder}
\end{equation}
where $\,\mathbf{e}_{x}$, $\,\mathbf{e}_{y}$ and $\,\mathbf{e}_{z}$ are the
space--fixed orthonormal Cartesian vectors (second set of axes). Notice that
we renamed the vector $\mathbf{r}_{2}^{\prime }$ in order to match Blinder's
notation\cite{B60}.

We next define the body--fixed orthonormal vectors
\begin{eqnarray}
\,\mathbf{e}_{x}^{\prime } &=&\frac{\,\mathbf{e}_{\theta }}{|\,\mathbf{e}%
_{\theta }|},\;\,\mathbf{e}_{\theta }=\frac{\partial \mathbf{R}}{\partial
\theta }  \nonumber \\
\mathbf{e}_{y}^{\prime } &=&\frac{\,\mathbf{e}_{\phi }}{|\,\mathbf{e}_{\phi
}|},\;\,\mathbf{e}_{\phi }=\frac{\partial \mathbf{R}}{\partial \phi }
\nonumber \\
\mathbf{e}_{z}^{\prime } &=&\,\mathbf{e}_{R},\;\,\mathbf{e}_{R}=\frac{%
\partial \mathbf{R}}{\partial R}  \label{eq:e'_Blinder}
\end{eqnarray}
and the new particle coordinates with respect to them
\begin{equation}
\mathbf{r}_{i}^{\prime }=x_{i}^{\prime }\mathbf{e}_{x}+y_{i}^{\prime }%
\mathbf{e}_{y}+z_{i}^{\prime }\,\mathbf{e}_{z}=x_{i}^{\prime \prime }\mathbf{%
e}_{x}^{\prime }+y_{i}^{\prime \prime }\mathbf{e}_{y}^{\prime
}+z_{i}^{\prime \prime }\mathbf{e}_{z}^{\prime }
\end{equation}
The transformation between these two sets of coordinates
\begin{equation}
\left(
\begin{array}{c}
x_{i}^{\prime \prime } \\
y_{i}^{\prime \prime } \\
z_{i}^{\prime \prime }
\end{array}
\right) =\mathbf{C}\left(
\begin{array}{c}
x_{i}^{\prime } \\
y_{i}^{\prime } \\
z_{i}^{\prime }
\end{array}
\right)  \label{eq:transf2_Blinder}
\end{equation}
is obviously given by the orthogonal matrix
\begin{equation}
\mathbf{C}=\left(
\begin{array}{ccc}
\mathbf{e}_{x}\cdot \mathbf{e}_{x}^{\prime } & \mathbf{e}_{y}\cdot \mathbf{e}%
_{x}^{\prime } & \mathbf{e}_{z}\cdot \mathbf{e}_{x}^{\prime } \\
\mathbf{e}_{x}\cdot \mathbf{e}_{y}^{\prime } & \mathbf{e}_{y}\cdot \mathbf{e}%
_{y}^{\prime } & \mathbf{e}_{z}\cdot \mathbf{e}_{y}^{\prime } \\
\mathbf{e}_{x}\cdot \mathbf{e}_{z}^{\prime } & \mathbf{e}_{y}\cdot \mathbf{e}%
_{z}^{\prime } & \mathbf{e}_{z}\cdot \mathbf{e}_{z}^{\prime }
\end{array}
\right) =\left(
\begin{array}{ccc}
\cos \phi \cos \theta & \sin \phi \cos \theta & -\sin \theta \\
-\sin \phi & \cos \phi & 0 \\
\cos \phi \sin \theta & \sin \phi \sin \theta & \cos \theta
\end{array}
\right)  \label{eq:C_Blinder}
\end{equation}
We do not show the explicit form of the Hamiltonian operator in this
molecule--fixed reference frame because it is not necessary for the present
discussion. Blinder\cite{B60} derived it for the HD molecule.

The new nuclear coordinates will be
\begin{eqnarray}
\mathbf{R}_{1} &=&\mathbf{r}_{1}-\frac{1}{2}(\mathbf{r}_{1}+\mathbf{r}_{2})=-%
\frac{\mathbf{R}}{2}=\left( 0,0,-\frac{R}{2}\right)  \nonumber \\
\mathbf{R}_{2} &=&\mathbf{r}_{2}-\frac{1}{2}(\mathbf{r}_{1}+\mathbf{r}_{2})=%
\frac{\mathbf{R}}{2}=\left( 0,0,\frac{R}{2}\right)
\end{eqnarray}
and the electronic ones are simply given by $\mathbf{r}_{i}^{\prime \prime }$%
, $i>2$. Assuming that both nuclei have identical charges then the classical
dipole moment in the body--fixed set of axes is given by a purely electronic
contribution
\begin{equation}
\mathbf{\mu }^{\prime \prime }=-e\sum_{i=3}^{N}\mathbf{r}_{i}^{\prime \prime
}  \label{eq:mu_body-fixed}
\end{equation}

For the particular case of HD ($N=4$) the expressions above agree with those
developed by Blinder\cite{B60}, except for the different notation and
labelling of nuclear and electronic coordinates. They allow one to calculate
the dipole moment as the expectation value of the corresponding operator by
means of an eigenstate of the Hamiltonian operator in the molecule--fixed
set of axes.

The effect of the inversion operator on the rotation angles is given by $%
\theta \rightarrow \pi -\theta $ and $\phi \rightarrow \phi +\pi $.
Therefore, the first row of the matrix $\mathbf{C}$ remains unchanged and
the other two ones change sign. For that reason the inversion operation $%
(x_{i}^{\prime },y_{i}^{\prime },z_{i}^{\prime })\rightarrow (-x_{i}^{\prime
},-y_{i}^{\prime },-z_{i}^{\prime })$ in the laboratory--fixed frame results
in $(x_{i}^{\prime \prime },y_{i}^{\prime \prime },z_{i}^{\prime \prime
})\rightarrow (-x_{i}^{\prime \prime },y_{i}^{\prime \prime },z_{i}^{\prime
\prime })$ in the body--fixed one, and, consequently, we do not expect that $%
\left\langle \hat{\mu}_{z}^{\prime \prime }\right\rangle $ vanishes because
of inversion symmetry. However, in the case of identical nuclei $m_{1}=m_{2}$
we cannot have a net dipole moment along the internuclear axis because the
additional permutational symmetry leads to $\left\langle \hat{\mu}%
_{z}^{\prime \prime }\right\rangle =0$. In order to explain the occurrence
of a dipole moment in a diatomic molecule of the form $^{m}$A$^{n}$A we
resort to perturbation theory and write $\hat{H}_{M}=\hat{H}_{0}+\lambda
\hat{H}^{\prime }$, where
\begin{equation}
\hat{H}_{0}=-\frac{\hbar ^{2}}{2m_{s}}\nabla _{2}^{\prime 2}-\frac{\hbar ^{2}%
}{2m_{e}}\left( \nabla _{3}^{\prime 2}+\nabla _{4}^{\prime 2}\right) -\frac{%
\hbar ^{2}}{8m_{s}}\left( \nabla _{3}^{\prime }+\nabla _{4}^{\prime }\right)
^{2}+V
\end{equation}
$\lambda =m_{s}/m_{a}$ and
\begin{equation}
\hat{H}^{\prime }=\frac{\hbar ^{2}}{2m_{s}}\nabla _{2}^{\prime }\cdot \left(
\nabla _{3}^{\prime }+\nabla _{4}^{\prime }\right)
\end{equation}
Notice that $\hat{H}_{0}$ is the molecular Hamiltonian for identical nuclei,
as follows from the fact that $\lambda =0$ when $m_{1}=m_{2}$.

We can thus expand the eigenfunction in a $\lambda $--power series $\Psi
=\Psi ^{(0)}+\Psi ^{(1)}\lambda \ldots $ that leads to a similar expansion
for the dipole moment $\mu _{z}=\mu _{z}^{(1)}\lambda +\ldots $ where
\begin{equation}
\mu _{z}^{(1)}=2\left\langle \Psi ^{(0)}\right| \hat{\mu}_{z}^{\prime \prime
}\left| \Psi ^{(1)}\right\rangle  \label{eq:mu_z^(1)}
\end{equation}
provided that the eigenfunction is chosen to be real. This simple argument
shows how the different nuclear masses produce a charge asymmetry and a net
dipole moment.

Blinder\cite{B60} estimated the dipole moment of HD by means of a rather
more complicated perturbation approach and later Kolos and Wolniewicz\cite
{KW63,KW66} and Wolniewicz\cite{W75,W76} improved Blinder's calculation by
means of a variational--perturbation method based on Eq.~\ref{eq:mu_z^(1)}.

It is clear that any rigorous nonadiabatic calculation of the molecular
dipole moment should be carried out in the molecule--fixed set of axes.
Consequently, those approaches outlined above in Sec.~\ref{sec:DM_NBO} are
unconvincing (to say the least). Cafiero and Adamowicz\cite{CA02a,CA02c}
must have placed the floating Gaussians in a convenient way to obtain
nonzero dipole moments in good agreement with the experimental ones and
Tachikawa and Osamura\cite{TO00} did not explain how they obtained their
results. Besides, the latter authors even forgot to remove the motion of the
center of mass.

\section{Perturbation theory for the Stark Shift}

\label{sec:PT} As we outline above in Sec.~\ref{sec:exp}, the experimental
determination of the molecular dipole moment relies on the validity of the
model of a rotating quasi--rigid polar body. The procedure consists of
fitting a polynomial function of the field strength to the observed
Stark--shift lines. In principle, we should derive more rigorous theoretical
expressions for those line shifts by means of perturbation theory and the
actual quantum--mechanical Hamiltonian operator for a molecule in an
electric field.

Instead of the rigid rotator model we should choose $\hat{H}_{0}=\hat{H}_{M}$
the Hamiltonian operator for the isolated molecule and the perturbation $%
\hat{H}^{\prime }=-\mathbf{F}.\mathbf{\hat{\mu}}$. If we set $\mathbf{F}$
along the $z$ axis in the laboratory frame ($\mathbf{F}=F\mathbf{e}_{z}$),
then $\hat{H}^{\prime }=-F\hat{\mu}_{z}$. In this way, perturbation theory
gives us the well--known quantum--mechanical expressions for the Stark
shifts
\begin{equation}
E_{n}=E_{n}^{(0)}+E_{n}^{(1)}+E_{n}^{(2)}+\ldots
\end{equation}
where $n$ is a collection of quantum numbers that completely specifies a
given (nonadiabatic) molecular state. For the ground state we have
\begin{eqnarray}
E_{0}^{(1)} &=&-\left\langle \Psi _{0}^{(0)}\right| \hat{\mu}_{z}\left| \Psi
_{0}^{(0)}\right\rangle F=0  \nonumber \\
E_{0}^{(2)} &=&F^{2}\sum_{m>0}\frac{\left| \left\langle \Psi
_{0}^{(0)}\right| \hat{\mu}_{z}\left| \Psi _{m}^{(0)}\right\rangle \right|
^{2}}{E_{0}^{(0)}-E_{m}^{(0)}}  \label{eq:E^(j)}
\end{eqnarray}
where we clearly appreciate that the field--reduced splitting $\Delta
E/F^{2} $ does not give us what we may call the square of the dipole moment
but a kind of energy--weighted average over those molecular states with
nonzero matrix element $\left\langle \Psi _{0}^{(0)}\right| \hat{\mu}%
_{z}\left| \Psi _{m}^{(0)}\right\rangle $, including the continuum part of
the spectrum. This important issue was already discussed by Brieger et al%
\cite{BRSH83} and Brieger\cite{B84} more than twenty years ago in their BO
study of the Stark effect of heteronuclear diatomic molecules in $^{1}\Sigma
$ states.

In the molecule--fixed reference frame we have
\begin{equation}
\mathbf{\mu }=\mu _{x}^{\prime }\mathbf{e}_{x}^{\prime }+\mu _{y}^{\prime }%
\mathbf{e}_{y}^{\prime }+\mu _{z}^{\prime }\mathbf{e}_{z}^{\prime }
\label{eq:mu_M}
\end{equation}
and a similar expression for $\mathbf{F}$. Brieger\cite{B84} argued that the
spherical vector components
\begin{eqnarray}
u_{0}^{\prime } &=&u_{z}^{\prime }  \nonumber \\
u_{\pm 1}^{\prime } &=&\mp \frac{1}{\sqrt{2}}(u_{x}^{\prime }\pm
u_{y}^{\prime })  \label{eq:sph_comp}
\end{eqnarray}
are more convenient for the calculation of the matrix elements in Eq.~(\ref
{eq:E^(j)}). It is not difficult to prove that
\begin{equation}
\mathbf{\mu }.\mathbf{F}=\sum_{p}(-1)^{p}\mu _{p}^{\prime }F_{p}^{\prime }
\label{eq:mu.F_sph}
\end{equation}
However, since the laboratory--fixed and molecule--fixed are the natural
frames for the external field and the molecule, respectively, then Brieger%
\cite{B84} chose $\mathbf{F}=F\mathbf{e}_{z}$ in the former and $\mathbf{\mu
}$ in the latter as in Eq.~(\ref{eq:mu_M}). Thus, in the simple notation of
Sec.~\ref{sec:Mol-Fixed}, we have
\begin{eqnarray}
\mu _{z} &=&\mu _{x}^{\prime }\mathbf{e}_{z}\cdot \mathbf{e}_{x}^{\prime
}+\mu _{y}^{\prime }\mathbf{e}_{z}\cdot \mathbf{e}_{y}^{\prime }+\mu
_{z}^{\prime }\mathbf{e}_{z}\cdot \mathbf{e}_{z}^{\prime }=  \nonumber \\
&=&-\sin \theta \,\mu _{x}^{\prime }+\cos \theta \,\mu _{z}^{\prime }=\frac{%
\sin \theta }{\sqrt{2}}\left( \mu _{+1}^{\prime }-\mu _{-1}^{\prime }\right)
+\cos \theta \,\mu _{0}^{\prime }  \label{eq:mu_z_M}
\end{eqnarray}

In this way, Brieger\cite{B84} showed that the perturbation $\hat{H}^{\prime
}$ connects the ground electronic state $^{1}\Sigma $ with excited $%
^{1}\Sigma $ and $^{1}\Pi $ ones (in general, those with electronic
angular--momentum quantum numbers $\Lambda $ and $\Lambda \pm 1$). Under the
BO approximation Brieger et al\cite{BRSH83} and Brieger\cite{B84} separated
the Stark shift (\ref{eq:E^(j)}) into four contributions:\ (I) coupling of
rotational states within the same $^{1}\Sigma $ vibronic one, (II) coupling
of vibrorotational states within the same $^{1}\Sigma $ electronic state,
(III) coupling of vibrorotational states between two $^{1}\Sigma $
electronic ones, and (IV) coupling of vibrorotational states between the
given $^{1}\Sigma $ and $^{1}\Pi $ ones. The first three contributions are
due to $\mu _{0}$ and the fourth one to $\mu _{\pm 1}$\cite{BRSH83,B84}.

As far as we know there is no nonadiabatic calculation of the Stark shift by
means of equation (\ref{eq:E^(j)}). A straightforward comparison of BO and
non BO calculations may lead to some difficulties as mentioned by Wolniewicz%
\cite{W76}: ``Since the familiar classification of electronic states of
diatomic molecules is based on the Born--Oppenheimer approximation, some
difficulties arise if one tries to use the standard nomenclature to describe
nonadiabatic functions''

\section{Conclusions}

\label{sec:conclusions} As said above the nonadiabatic calculations of
dipole moments are scarce and few. Those of Tachikawa and Osamura\cite{TO00}
and Cafiero and Adamowicz\cite{CA02a,CA02c} are not reliable because they
are not based on the molecule--fixed Hamiltonian operator. In our opinion
the only serious attempts to the nonadiabatic calculation of dipole moments
are those of Blinder\cite{B60,B61}, Kolos and Wolniewicz\cite{KW63,KW66} and
Wolniewicz\cite{W75,W76}. All the other calculations of dipole moments are
based on the BO approximation.

We have also seen that the comparison between theoretical and experimental
dipole moments is not straightforward. The field--reduced line splittings do
not give us the square of the molecular dipole moment, as predicted by the
oversimplified rigid--rotor model, but a kind of energy--weighted average.
Therefore, there is much to be done in this field and we hope that the
present discussion will contribute to motivate such work.

\begin{table}[tbp]
\caption{Dipole moments for some diatomic molecules}
\label{tab:DM}
\begin{center}
\begin{tabular}{lll}
\hline
Ref. & $\mu \ (D)$ & Method \\ \hline
\multicolumn{3}{c}{LiH} \\ \hline
\cite{CA02a} & 5.8816 & non BO \\
\cite{PWHU96} & 5.879 & BO \\
\cite{TO00} & 6.072 & non BO \\
\cite{LT74} & 5.8820 (4) & Exp. \\
\cite{WGK62} & $5.882 \pm 0.003$ & Exp. \\ \hline
\multicolumn{3}{c}{LiD} \\ \hline
\cite{CA02a} & 5.8684 & non BO \\
\cite{TO00} & 6.080 & non BO \\
\cite{LT74} & 5.8677 (5) & Exp. \\
\cite{WGK62} & $5.868 \pm 0.003$ & Exp. \\ \hline
\multicolumn{3}{c}{HD} \\ \hline
\cite{KW66} & $1.54 \times 10^{-3}$ & non BO \\
\cite{CA02c} & $0.831 \times 10^{-3}$ & non BO \\ \hline
\multicolumn{3}{c}{HT} \\ \hline
\cite{CA02c} & $1.111 \times 10^{-3}$ & non BO \\ \hline
\multicolumn{3}{c}{DT} \\ \hline
\cite{CA02c} & $2.77 \times 10^{-4}$ & non BO \\ \hline
\end{tabular}
\end{center}
\end{table}

\end{document}